\documentclass[
showpacs,
bibnotes,
amsmath,amssymb,
pra,
twocolumn,                   %Modify
letterpaper,
twoside,
frontmatterverbose,
floatfix,
notitlepage,
]{revtex4-1}

\usepackage{graphicx}% Include figure files
\usepackage{dcolumn}% Align table columns on decimal point
\usepackage{bm}% bold math
\usepackage{hyperref}
\usepackage{multirow}
\usepackage{array}
\usepackage{cancel}
\usepackage{booktabs}
\usepackage{ctable}
\usepackage{upgreek}
\usepackage{epsfig}
\usepackage{mathrsfs}
\usepackage{amssymb}
\usepackage{amsbsy}
\usepackage{color}
\usepackage{cancel}
\usepackage[normalem]{ulem}
\usepackage{pifont}
\usepackage{marginnote}
\usepackage{float}
\usepackage{tikz}
\usetikzlibrary{arrows}
\tikzstyle{block}=[draw opacity=0.7,line width=1.4cm]
\usetikzlibrary{positioning}
\usepackage[caption=false]{subfig}
\usepackage{setspace}
\newcommand{\bcen}{\begin{center}}
\newcommand{\ecen}{\end{center}}
\newcommand{\btab}{\begin{tabular}}
\newcommand{\etab}{\end{tabular}}
\newcommand{\bdes}{\begin{description}}
\newcommand{\edes}{\end{description}}

\newcommand{\beq}{\begin{equation}}
\newcommand{\eeq}{\end{equation}}
\newcommand{\bea}{\begin{eqnarray}}
\newcommand{\eea}{\end{eqnarray}}

\newcommand{\bary}{\begin{array}}
\newcommand{\eary}{\end{array}}
\newcommand{\benum}{\begin{enumerate}}
\newcommand{\eenum}{\end{enumerate}}
\newcommand{\bitem}{\begin{itemize}}
\newcommand{\eitem}{\end{itemize}}

\makeatletter

\newcommand{\Rmnum}[1]{\expandafter\@slowromancap\romannumeral #1@}
\makeatother

\newcommand{\kfo}{{k^F_{0}}}
\newcommand{\hopt}{{t_{hop}}}

\newcommand{\authorprl}[2]{\author{#1}\email{#2}}

\newcommand{\titlename}{Boosted one dimensional superfluids on a lattice}

\begin{document}

%\preprint{}
%\preprint{}


\title{\titlename}
\authorprl{Sayonee Ray}{sayoneeray@physics.iisc.ernet.in}
\authorprl{Subroto Mukerjee}{smukerjee@physics.iisc.ernet.in}
\authorprl{Vijay B. Shenoy}{shenoy@physics.iisc.ernet.in}
\affiliation{Centre for Condensed Matter Theory, Department of Physics, Indian Institute of Science, Bangalore 560 012, India}
\date{\today}


\makeatletter
\let\savetitle\@title
\let\saveauthor\@author
\makeatother 

\begin{abstract}
We study the effect of a boost (Fermi sea displaced by a finite momentum) on one dimensional systems of lattice fermions with short-ranged interactions. In the absence of a boost such systems with attractive interactions possess algebraic superconducting order. Motivated by physics in higher dimensions, one might naively expect a boost to weaken and ultimately destroy superconductivity. However, we show that for one dimensional systems the effect of the boost can be to strengthen the algebraic superconducting order by making correlation functions fall off more slowly with distance. This phenomenon can manifest in interesting ways, for example, a boost can produce a Luther-Emery phase in a system with both charge and spin gaps by engendering the destruction of the former.
\end{abstract}

\maketitle

\paragraph*{Introduction:}  An electrical current set up in a superconductor continues to flow even in the absence of a driving electric field~\cite{santos,bagwell,tinkham}. Such a persistent current is equivalent to an imbalance in the number of carriers moving along and opposite to the direction of the current, i.e. a boost. In a 1D system, a boost can be realized by having different chemical potentials for left and right movers. The magnitude of the boost or the current cannot be arbitrarily large and there is a critical value above which the superconducting state is destroyed. This phenomenon is analogous to the destruction of a superfluid when its flow velocity is larger than the critical velocity. The critical value of the boost can be calculated from the Bogoliubov- de Gennes equations for a superfluid~\cite{baym} and superconductor~\cite{wei,zagoskin}.
%\sout{What is this $u$}.
%\cancel{Also this.}

A natural question to ask is about the fate of superconductors, which do not have long range order (and hence order parameter equal to zero) upon the application of a boost. The most common example of such a system is a one dimensional system of fermions with attractive interactions~\cite{giamarchi,gnt,schulz,haldane}. Such one dimensional superconductors have recently come to the fore as they possess interesting topological properties such as the existence of Majorana edge modes under appropriate conditions~\cite{kitaev,nayak,kraus}. Experiments to detect these modes typically involve driving a current through the superconductor \cite{yazdani,shach} and hence it is germane to ask how large the critical current in these systems can be. Moreover, such systems have also been realized in cold atomic gases where it has been possible to make the system left-right asymmetric thereby producing a boost~\cite{zwierlein,wang}. A boosted clean one dimensional superconductor has been studied within mean-field theory and its critical velocity calculated~\cite{wei}. It was found that the standard Landau critical velocity is replaced by a smaller value due to a pre-emptive Clogston-Chandrasekhar-type discontinuous transition. However, a similar calculation for a clean one dimensional superconductor incorporating the effects of quantum fluctuations has not been performed so far. However, it has been shown that phase slips induced by the contact of the superconductor with the walls of a container or the presence of statically irrelevant perturbations can dynamically destroy superconductivity at finite frequency and temperature in one dimension~\cite{oshikawa}.

In this paper, we address and answer the question of how a boost affects the quasi-long-ranged superconducting state in one dimension. Our main result is that the boost can have the counter-intuitive effect of {\em strengthening} the superconductivity (in a sense that we explain later) as opposed to weakening it like in higher dimensions. The boost eventually destroys superconductivity at a critical value but does so discontinuously when one of the Fermi points of the system is boosted to zero momentum. We demonstrate that a similar effect exists even for systems with quasi-long-ranged charge density wave order, i.e, the order is strengthened upon the application of a boost. We also show that for the boost to have any non-trivial effect, the underlying system has to have broken Galilean invariance, which is naturally realized in lattice systems. This has the additional effect of producing interesting phases at commensurate filling upon the application of a boost when umklapp is operative. These observations point to the possibility of new types of phase transitions that can be achieved by boosting the system. For example, we show that a system with a charge and spin gap can be boosted into a Luther-Emery phase \cite{le1,le2} by closing the charge gap. 
%The realization of the Luther-Emery phase in cold atomic systems is currently an active area of research \cite{le2} and our calculations suggest a new way of doing this.}

\paragraph*{Framework:} 

\begin{figure}%[H]
\centering
\includegraphics[width=0.95\columnwidth]{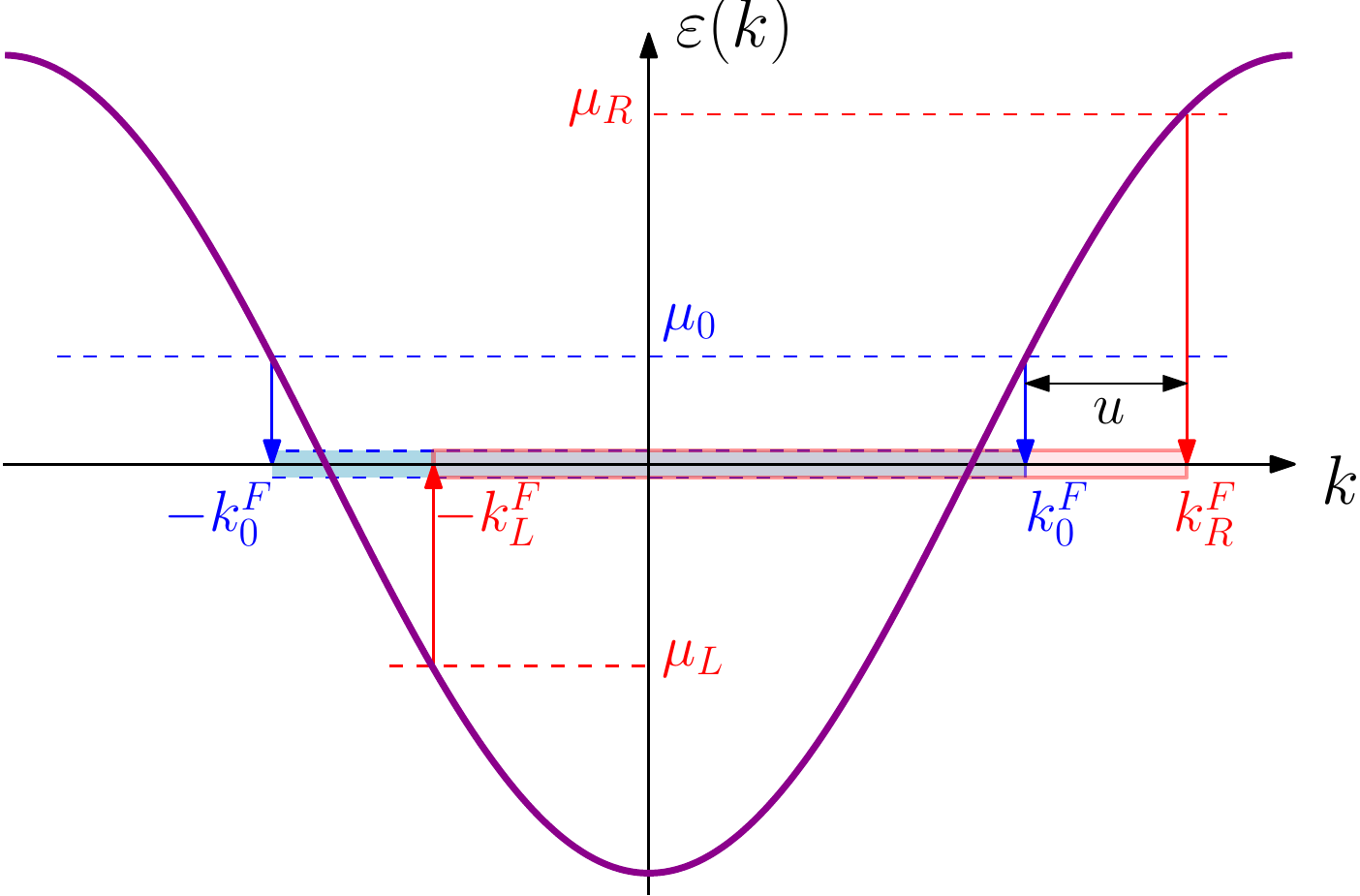} 
\caption{Schematic diagram of the boosted Fermi sea. $\kfo$ is the unboosted Fermi momentum with the chemical potential $\mu_0$. Upon boosting the chemical potential of the left movers is $\mu_L$ while that of the right movers is $\mu_R$ with concomitant Fermi wavevector of $k^F_L$ and $k^F_R$, both of which depend on the boost $u$, via Eq. (\ref{Eq:kFu}). }
\label{fig:scheme}
\end{figure}

One dimensional systems of spinless fermions with a dispersion $\epsilon(k)$ symmetric in $k$ have two Fermi points at $\kfo$ and $-\kfo$ corresponding to right and left movers respectively. $\kfo=\pi n$, where $n$ is the density of fermions. The boost $u$ is defined by a transformation $k \rightarrow k+u,~\forall k$. This destroys the left-right symmetry and the two Fermi points are now at,
\begin{equation} \label{Eq:kFu}
k ^F_s=\kfo +s u, 
\end{equation}
where $s=1[-1]$ for the right[left] movers with Fermi velocity $v^F_s (u)= \left. \frac{d \epsilon}{dk} \right|_{k=k^F_s(u)}.$ Assuming $\epsilon(k)$ to be analytic, 
$$v^F_s=v^F(u)+s w(u),$$ 
where $v^F(u)$ and $w(u)$ are even and odd functions of $u$ respectively.  The effective low energy theory in the presence of interactions can be obtained by bosonizing about these boosted Fermi points producing a Hamiltonian  
 
\begin{equation} \label{Eq:Ham_harm}
\begin{split}
\mathcal{H} = & \frac{v(u)}{2}\int_{-L/2}^{L/2} dx \bigg[K(u) :\left(\Pi(x) \right)^2: + \frac{1}{K(u)} :\left(\nabla \phi(x) \right)^2: \\
&- 2 w(u):\left( \nabla \phi(x) \Pi(x) \right):\bigg],
\end{split}
\end{equation} 

where $\phi$ and $\Pi=\partial_t \phi$ are the bosonic field and its conjugate respectively and $: ():$ denotes normal ordering \cite{giamarchi} (and see the supplemental material~\cite{supplemental}). Further, 
\begin{eqnarray}\nonumber
v(u) & = & \sqrt{\left(v^F(u) + \frac{g_4}{2\pi} \right)^2 - \left( \frac{g_2}{2\pi} \right)^2},\\ 
K(u)& = & \sqrt{\frac{v^F(u) - \left(\frac{g_2}{2\pi}-\frac{g_4}{2\pi}\right)}{v^F(u) + \left(\frac{g_2}{2\pi}+\frac{g_4}{2\pi}\right)}},
\label{Eq:boostpara}
\end{eqnarray} 
are the effective Fermi velocity and Luttinger parameter respectively. $g_4$ [$g_2$] is the strength of the scattering of fermions on the same [opposite] side of the 1D Fermi sea
$g_4$ and $g_2$ remain unchanged even for $u \neq 0$ since the amplitudes for the scattering processes only involve the momentum difference of the participating particles (See the supplemental material~\cite{supplemental}). Eq. (\ref{Eq:boostpara}) shows $v(u)$ and $K(u)$ pick up $u$ dependences only from $v^F(u)$ and $u$ also introduce a coupling between $\Pi$ and $\partial_x \phi$ with strength $w(u)$. It can also be seen that Eqs. (~\ref{Eq:Ham_harm}) and (~\ref{Eq:boostpara}) reduce to their standard forms when $u=0$~\cite{giamarchi,gnt,schulz,haldane}. When $u=0$, the system possesses conformal invariance since the coupling between the $\Pi$ and $\phi$ fields is absent, which the boost clearly breaks. However, conformal invariance can be restored by introducing new fields:
\begin{eqnarray}
\tilde{\phi}(x,t) &=& \phi[x+w(u)t,t] \\ 
\tilde{\theta}(x,t) &=& \theta[x+w(u)t,t],
\end{eqnarray}
in terms of which the Hamiltonian can again be written in the standard form with Fermi velocity $v(u)$ and Luttinger parameter $K(u)$ :
\begin{equation}
\mathcal{H}= \frac{v(u)}{2}\int_{-L/2}^{L/2} dx \left[K(u) :\left(\tilde{\Pi}(x) \right)^2: + \frac{1}{K(u)} :\left(\nabla \tilde{\phi}(x) \right)^2: \right],
\label{Eq:Ham_tilde}
\end{equation}
Thus, conformal invariance can be restored by effecting a (non-conformal) Galilean transformation on the space-time co-ordinates. $w(u)$ acts as the velocity in this transformation.
For small $u$ (compared to $\kfo$), 
\begin{equation}
K(u) \approx K(0) + \frac{u^2}{2} \left. \frac{dK(u)}{dv^F(u)} \frac{d^2 v^F(u)}{du^2} \right|_{u=0}.
\label{Eq:small_boost}
\end{equation}
 It can be seen that for a system with Galilean invariance (and hence $\epsilon(k)$ quadratic in $k$), $K(u)=K$. Thus, Galilean invariance needs to be broken for a non-trivial effect of the boost for which we put the fermions on a lattice with nearest neighbour hopping $- \hopt$ and dispersion $\epsilon(k)=-2 \hopt \cos k$. 
Consequently, 
\begin{eqnarray}\nonumber
v^F(u) & = & v^F(0)\cos u \\
w(u) & = & 2 \hopt \cos (\kfo) \sin u,
\label{Eq:lattice}
\end{eqnarray} 
where $v^F(0)= 2 \hopt \sin (\kfo)$ and thus
\begin{equation}
K(u) \approx K(0) +\frac{u^2}{4} \left[ \frac{v^F(0)}{v(0)} \left(K(0)^2-1 \right)\right].
\label{Eq:Kwithu}
\end{equation}

Eq. (\ref{Eq:Kwithu}) implies that $K(u) > K(0) [K(u) < K(0)]$ for $K(0) >1 [K(0)<1]$ for small $u$. This is true at larger values of $u$ as well. Further, it can be seen that $K(u)$ depends not just on $K(0)$ and $u$ but also the microscopic parameters $v^F(0)$, $g_2$ and $g_4$ (the latter two through $v(0)$), showing that the way the boost modifies the Luttinger parameter is "not universal".  

It is known that $K>1$ and $K<1$ result in dominant quasi-long-range ordered superconducting (SU) and charge density wave (CDW) order respectively~\cite{giamarchi,gnt,schulz,haldane}. This continues to be true even for $u \neq 0$ since the system is described by the standard harmonic Hamiltonian under the transformations $K \rightarrow K(u)$, $v^F \rightarrow v^F(u)$ and $\phi \rightarrow \tilde{\phi}$ but in terms of the transformed space-time coordinates. The SU and CDW correlation functions are thus given by\cite{giamarchi,miranda}
\begin{widetext} 
\begin{eqnarray}\nonumber 
\langle O_{\rm SU}(x,t) O_{\rm SU}^\dagger (x',t') \rangle & \sim & e^{i 2 u(x-x')} \left( \frac{1}{\sqrt{\left[x-x' + w(u)(t-t')\right]^2+\left[v^F(u) \right]^2 (t-t')^2}} \right)^{1/K(u)}, \\
\langle O_{\rm CDW}(x,t) O_{\rm CDW} (x',t') \rangle & \sim & \cos{[2 \kfo(x-x')]} \left( \frac{1}{\sqrt{\left[x-x' + w(u)(t-t')\right]^2+\left[v^F(u) \right]^2 (t-t')^2}} \right)^{K(u)},
\label{Eq:6}
\end{eqnarray}
\end{widetext}
where $O_{\rm SU[CDW]}$ is the SU[CDW] order parameter (See the supplemental material~\cite{supplemental}). $\langle O_{\rm SU[CDW]}(x,t) \rangle =0$ since there is no long range order. The loss of conformal invariance upon the application of a boost can be clearly seen from the asymmetric way in which the space and time coordinates appear in Eq. (\ref{Eq:6}). The equal time correlation functions can be obtained by setting $t=t'$ and it can be seen that the SU[CDW] order decays algebraically with distance with exponent $1/K(u) [K(u)]$. Thus, the order that was dominant in the absence of a boost is strengthened by it while the sub-dominant one is weakened. {\em Hence, a system that is superconducting has its superconductivity strengthened in the presence of a boost.} A CDW system also has its CDW order strengthened similarly. 

Note, however, that this is true only for $u < \kfo$ since beyond that value, there is only one species (either left or right moving) of fermions. Superconductivity is thus discontinuously destroyed a this critical value of the boost. Suppose $\kfo < \pi/2$, one of the Fermi points will move to $k=0$ when a boost $u=\kfo$ is applied. It will no longer be possible to linearize about this Fermi point and so $u=\kfo$ might be a natural limit to the applicability of our treatment. Even though, it once again becomes possible to linearize when the Fermi point moves to $k > 0$, the Fermi vacuum becomes unstable and hence the system is not amenable to our treatment\cite{Note1}. %%%%%\footnote{If both the Fermi points $k^F_L$ and $k^F_R$ are positive, then linearization of the dispersion at $k^F_L$ becomes problematic since the unoccupied states $(k < k^F_L)$ have lower energy rendering the Fermi sea unstable.}.

A simple way of understanding this physics is by noting that the boost produces an enhancement of the ``pairing susceptibility'' by enhancing the particle-particle density of states (See supplemental information~\cite{supplemental}). This point also suggests that the enhancement is not universal, and depends on the details of the dispersion. For example, for the usual tight binding model $ v(u)$ decreases with $u$ and $w(u)$ is positive (for $u>0$). However, this is not generic, one can have 1D systems where $w(u)$ is negative and $v(u)$ increases with $u$. In fact a generic 1D lattice dispersion can have varying effects based on the position of $k_F$, the local nature of the dispersion can affect the strengthening or weakening of the long-range superconducting correlations.

\paragraph*{Effect of spin and umklapp:}  Having analyzed the effect of a boost on a system of spinless fermions, we now turn our attention to spin $1/2$ systems. In the absence of a boost, it is known that the charge and spin degrees of freedom can be separated in the low-energy physics, each being described by its own hamiltonian $H_\nu$, fields $\Pi_\nu$ and $\phi_\nu$, Fermi velocity $v_\nu$ and Luttinger parameter $K_\nu$, where $\nu=\rho [\sigma]$ for the charge [spin] sector~\cite{giamarchi}. A point of difference between the sectors is that the spin sector has umklapp even when the underlying system possesses Galilean invariance while the charge sector does not. Umklapp can be relevant in the charge sector only for systems with broken Galilean invariance and commensurate filling. Since we need to break Galilean invariance for the boost to have a non-trivial effect, the low energy physics of the spin and charge sectors is described by the Hamiltonian in Eq. (\ref{Eq:Ham_umklapp}).

\begin{widetext}
 \begin{equation}
\mathcal{H_\nu}= \frac{v_\nu (u)}{2}\int_{-L/2}^{L/2} dx \left[K_\nu (u):\left(\tilde{\Pi}_\nu(x) \right)^2: + \frac{1}{K_\nu (u)} :\left(\partial_x \tilde{\phi}_\nu (x) \right)^2:  + \frac{g_\nu }{a^2} :\cos(\alpha_\nu  \tilde{\phi}_\nu ): \right],
\label{Eq:Ham_umklapp}
\end{equation}
\end{widetext}
where $g_\nu$ is the strength of the umklapp term, $a$ an ultra-violet cutoff and $\alpha_\rho= \sqrt{16 \pi}$ and $\alpha_\sigma= \sqrt{8 \pi}$  for charge and spin respectively. We emphasize again that $g_\rho$ is operative only at commensurate filling although for a system in contact with a container, a similar term may arise with a phase oscillating is space with a minimum wavenumber~\cite{oshikawa}.
Again, the Hamiltonian reduces to the standard form~\cite{giamarchi,gnt,schulz,haldane} when $u=0$. 
 
The renormalization group flow equations for the parameters $g_\nu$ and $K_\nu$ at tree level, in terms of two new parameters $h_\nu=2\left( \frac{K_\nu}{K^c_\nu}-1 \right)$ and $g^\perp_\nu=K^c_\nu g_\nu$ with $K^c_\nu = \frac{8 \pi}{\alpha_\nu^2}$ are:  \begin{eqnarray} \label{Eq:RG_flow}
\frac{d g^\perp_\nu}{d l}\ &=\ - h_\nu g^\perp_\nu,\\ \nonumber
  \frac{d h_\nu}{d l}\ &=\ - \left( g^\perp_\nu \right)^2.
%\label{Eq:RG_flow}
\end{eqnarray}
The above equations can be integrated to obtain flow lines and for $h_\nu<0$, $g^\perp_\nu$ is a relevant perturbation and opens a gap. For $h_\nu>0$, the flow terminates at $g^\perp_\nu=0$ and $h_\nu=h^*_\nu$ (i.e. a Luttinger liquid results). This has the value $h_{\nu,i}^2 - \left( g^\perp_{\nu,i} \right)^2 = \left(h^*_\nu \right)^2$, where $h_{\nu,i}$ and $g^\perp_{\nu,i}$ are the initial (bare) values of $h_\nu$ and $g^\perp_\nu$.

Eqs. (~\ref{Eq:RG_flow}) are valid even for $u\neq 0$. The only effect of the boost is to change the values of the initial parameters in the following way (See the supplemental material~\cite{supplemental})
\begin{eqnarray} \nonumber
h_{\nu,i}(u) & = & 2 \bigg(\frac{K_{\nu,i}(0)}{K^c_\nu}-1 \bigg) - \frac{v^F_{i}(0)  f(u)}{v_{\nu}(0) K^c_\nu}  \left[\left(K_{\nu,i}(0) \right)^2 - 1\right]  \\ 
g^\perp_{\nu,i}(u)& = & g^\perp_{\nu,i}(0) \left[1 - \frac{v^F_{i}(0) f(u)}{2 v_{\nu}(0)} \bigg(K_{\nu,i}(0) + \frac{1}{K_{\nu,i}(0)} \bigg) \right].
\label{Eq:hg_boost}
\end{eqnarray}
where $f(u) = - \frac{v^F_{i}(0)-v^F_{i}(u)}{v^F_{i}(0)},$ is assumed to be small and $K_{\nu,i}(0) = K^c_\nu [1+h_{\nu,i}(0)/2]$. Note that the first of the above equations is the same as Eq. (\ref{Eq:Kwithu}) but with $K^c_\rho=1/2$ as is appropriate for spinless fermions. Even in this more general case with spin, the boost has the effect that $K_{\nu,i} (u) > K_{\nu,i}(0) [K_{\nu,i} (u) <  K_{\nu,i}(0)]$ if $K_{\nu,i}(0) > 1 [K_{\nu,i}(0) < 1]$.

The value $h_\nu^*(u)$ can be determined for a flow staring at $h_{\nu,i}(u)$ and $g^\perp_{\nu,i}(u)$ and is given by: %\begin{widetext}
\begin{equation}
\begin{split}
\left[h_\nu^*(u)\right]^2 &= 4 \left(\frac{K_{\nu,i}(0)}{K^c_\nu} - 1 \right)^2 -\left(g^\perp_{\nu,i}(0) \right)^2 \\
& + \frac{v^F_{i}(0) f(u) }{v_{\nu}(0) K^c_\nu} \bigg[4 (K_{\nu,i}(0)^2 - 1) \left(1 - \frac{K_{\nu,i}(0)}{K^c_{\nu,i}} \right) \\
& + \frac{g^\perp_{\nu,i}(0)^2 K^c_\nu}{K_{\nu,i}(0)} (K_{\nu,i}(0)^2 + 1) \bigg].
\end{split}
\end{equation}
%\end{widetext}

It can be seen that in the presence of umklapp a sector is gapped (gapless) when $K_\nu^*(u) < K^c_\nu \left[ K_\nu^*(u) \geq K^c_\nu \right]$. For spinless fermions, $K^c_\rho=1/2 < 1$ ~\cite{giamarchi,gnt} and so when superconductivity dominates in the charge sector ($K_{\rho, i}(0)>1$), it is strengthened when the system is boosted just like in the absence of umklapp and the charge sector continues to be gapless. When $K_{\rho,i}(0) < 1$, gapless CDW order results down to a critical value of $g^\perp_{\rho,i}$, below which a gapped state is obtained, which can even result in a long range CDW case. 

\begin{figure}
%\centering 
%\begin{tabular}{c} 
\includegraphics[width=0.95\columnwidth]{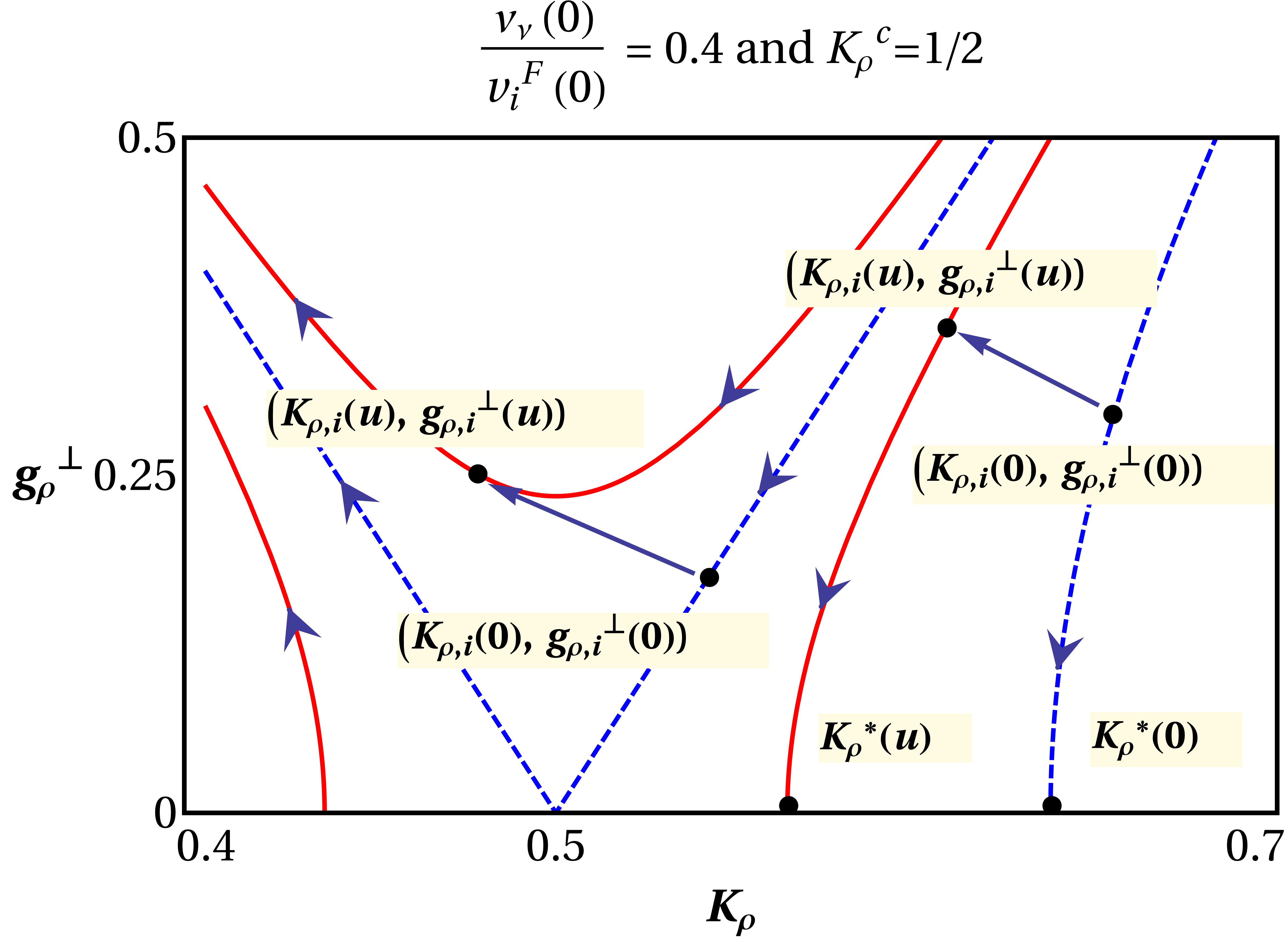} %\\

\includegraphics[width=0.95\columnwidth]{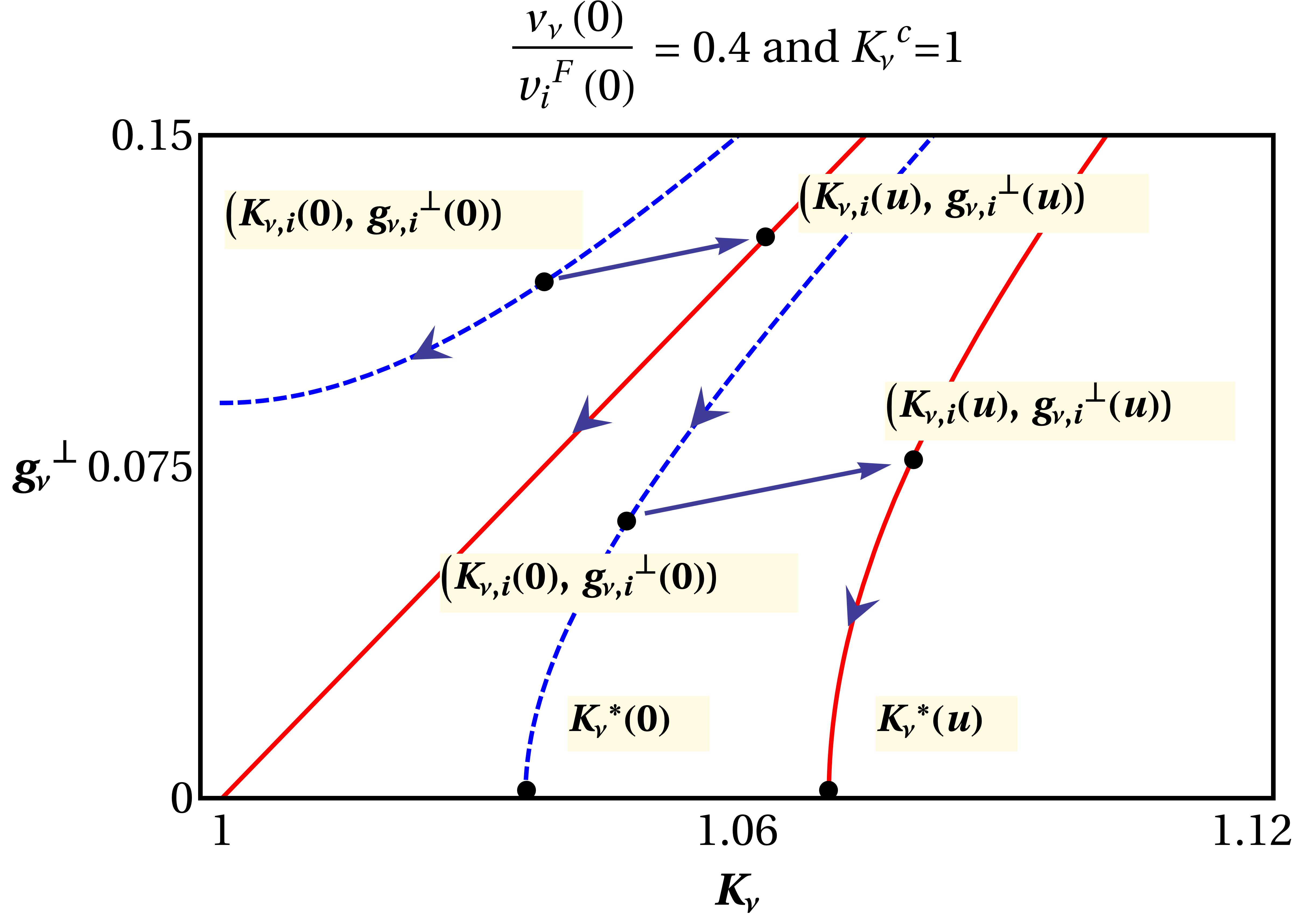}
%\end{tabular}
\caption{The effect of a boost on the renormalization group flows of the parameters $K_\nu$ and $g^\perp_\nu$. The boost changes the initial values $K_{\nu,i}$ and $g^\perp_{\nu,i}$ after which the flows are given by Eqs. (~\ref{Eq:RG_flow}). This is shown for a system of spinless fermions (top panel) with two loci of initial conditions (dashed blue lines) which under the effect of a boost are mapped onto two different loci of initial conditions (red solid lines). The value of $K^*_\rho$ changes as a result. Additionally, the boost can also transform a locus of initial values flowing to a gapless phase into one which flows to a gapped phase as shown. The critical value of $K_\rho$ separating the gapped and gapless phases is $K^c_\rho=1/2$. The same effect on a system of spin 1/2 fermions (bottom panel). Here, there are two decoupled sectors corresponding to $\nu=\rho,\sigma$ and $K^c_\nu=1$. This has the effect that a boost can no longer open a gap: instead it can transform a locus of points flowing to a gapped phase into one which flows to a gapless phase.
\label{fig:RGflow}}
\end{figure}

It can be seen from Fig.~\ref{fig:RGflow} that upon the application of a boost (which has the effect of reducing the value of $K_{\rho,i}$), a gapless CDW state can be transformed into a gapped one. {\em Thus a boost can convert quasi-long-ranged CDW order into true long-range order.} If the CDW state continues to remain gapless upon the application of a boost, the order is strengthened like in the case without umklapp.

\begin{figure}
%\begin{tabular}{c}
\centerline{\includegraphics[width=0.5\columnwidth]{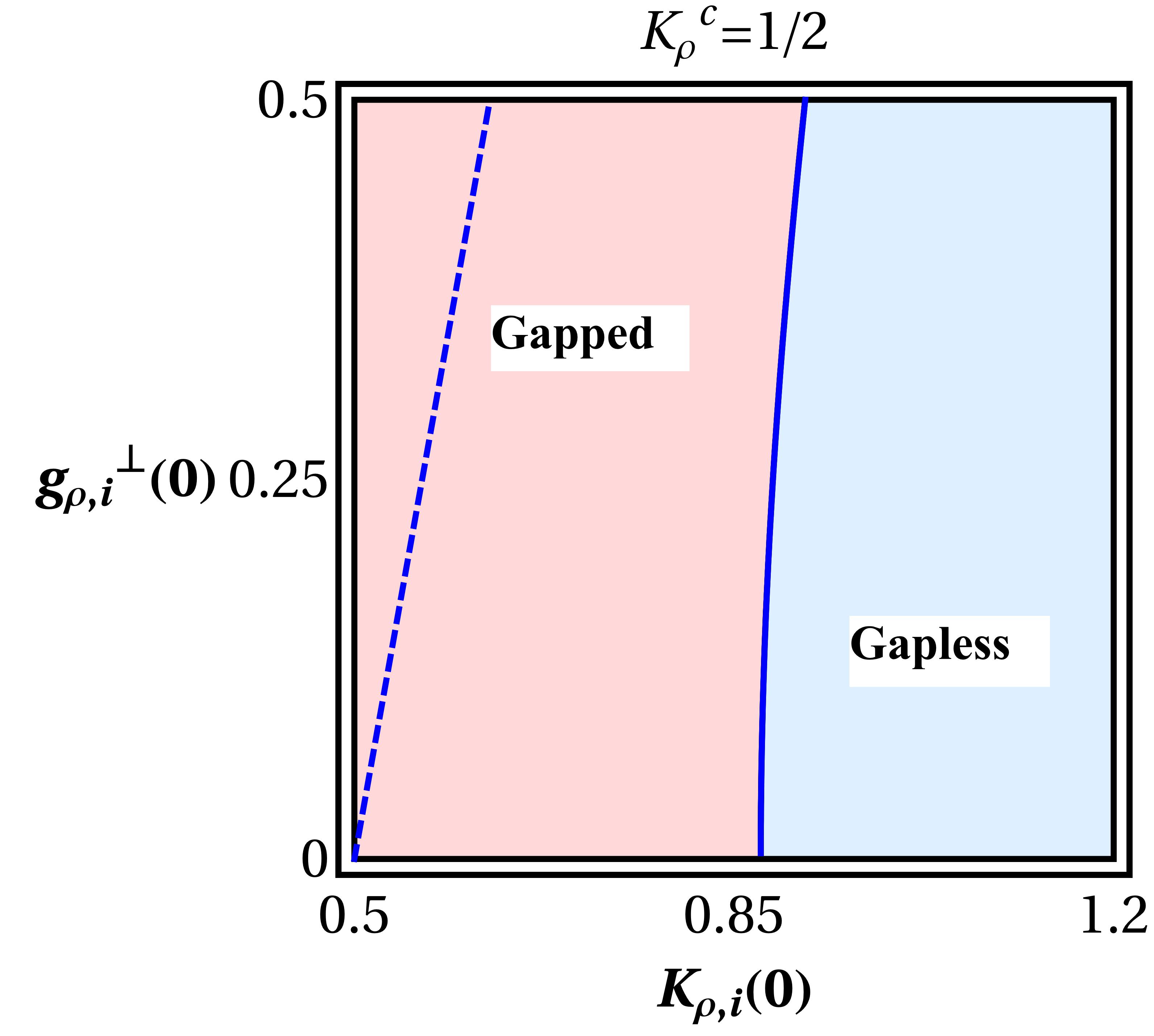}~
\includegraphics[width=0.5\columnwidth]{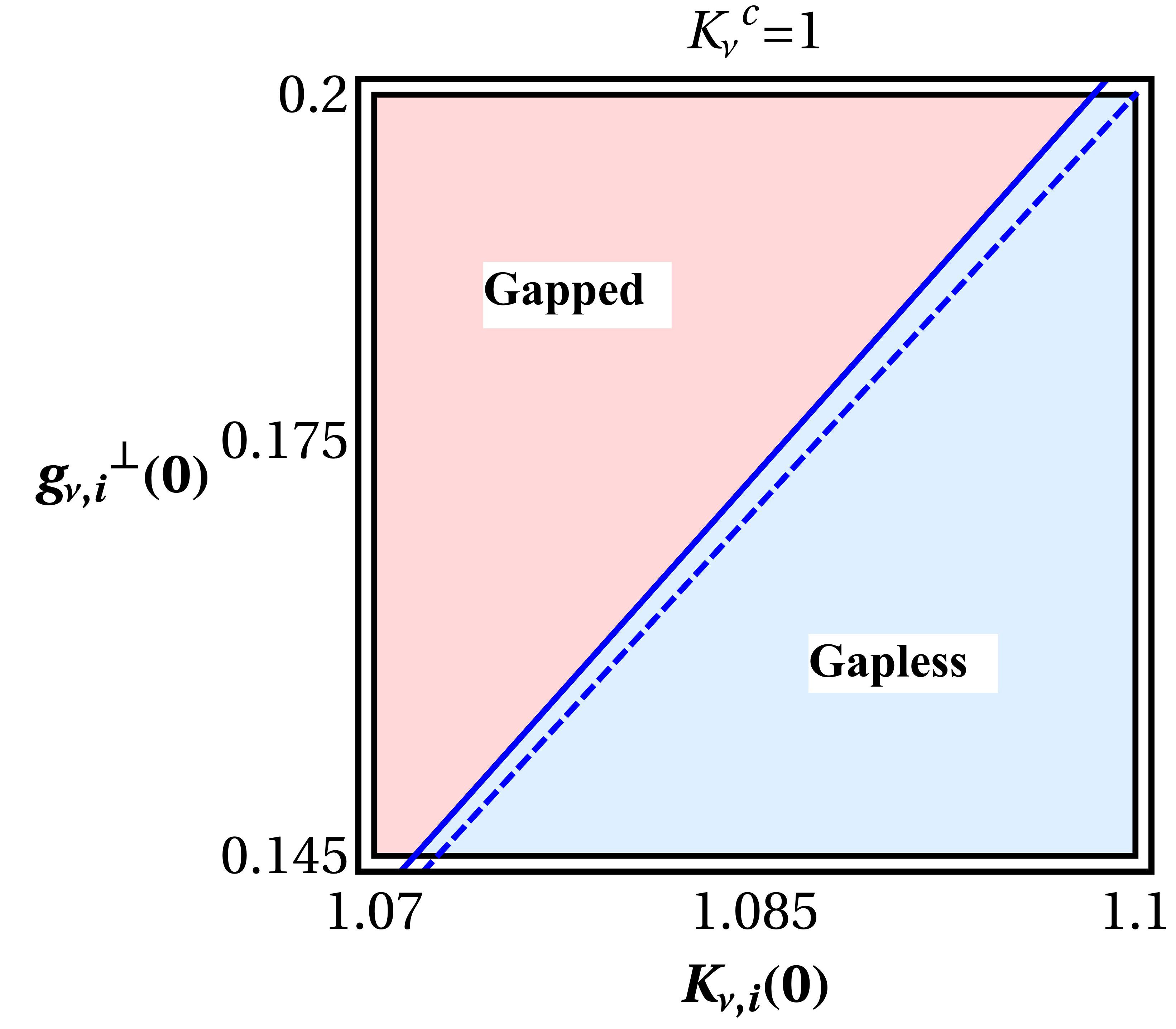}}
%\end{tabular}
\caption {The transformation that can be effected by a boost on a system of spinless (left panel) and spin 1/2 (right panel) fermions. For spinless fermions, the region to the right [left] of the dashed line is gapless [gapped] without a boost. The region between the dashed and solid lines can become gapped upon the application of a suitably large boost. The region to the left [right] of the dashed [solid] line continues to be gapped [gapless]. For spin 1/2 fermions, the opposite happens and a boost closes an existing gap. The region to the right [left] of the solid line is gapless [gapped] without a boost. The region between the solid and dashed lines can become gapless upon the application of a suitably large boost. The region to the left [right] of the solid [dashed] line continues to be gapped [gapless].  
\label{fig:PhaseDiagram}}
\end{figure}

For spinful fermions on a lattice, $K^c_\rho=K^c_\sigma=1$~\cite{giamarchi}. Consequently, for $K_{\rho,i} < 1$, the system is always gapped and it is possible to have such a phase even when $K_{\rho,i} \geq 1$ depending on the value of $g^\perp_{\rho,i}$ as can be seen in Fig.~\ref{fig:RGflow}. A boost cannot open a charge gap in this case unlike for spinless fermions. However, it can close an existing gap for systems with a certain range of values of $K_{\rho, i}$ and $g^\perp_{\rho, i}$ as can be seen in Fig.~\ref{fig:PhaseDiagram}. This happens only for $K_{\rho,i}>1$. The boost has exactly the same effect in the spin sector as well.
 
The above conclusions open up the possibility of scenarios in which a system with a gap in the charge or spin sector or both can be transformed into a different phase by closing one or both gaps upon the application of a boost. Of particular interest is the situation where the system has both a charge and spin gap. If $K_{\rho,i}(0)$ and $g^\perp_{\rho,i}(0)$ lie in the red colored region between the dashed and the solid line in Fig.~\ref{fig:PhaseDiagram}, a boost can open up a charge gap. If $K_{\sigma, i}(0)$ and $g^\perp_{\sigma,i}(0)$ lie in the red region (beyond the solid line), the boost cannot close the spin gap and the resultant state is a Luther-Emery fluid with gapped spin excitations and gapless charge excitations. {\em Thus, it is possible to obtain a Luther-Emery fluid from a fully gapped system by applying a boost which suggests a new way of obtaining such a fluid in experiments on trapped cold atoms ~\cite{zwierlein,wang}}. It is also possible to destroy the spin gap of a Luther-Emery fluid by applying a boost if $K_{\sigma, i}(0)$ and $g^\perp_{\sigma,i}(0)$ for the systems lie in the blue colored region between the solid and the dashed lines, as shown in Fig.~\ref{fig:PhaseDiagram}.

To conclude, we have shown that the application of a boost can strengthen the superconductivity of typical one dimensional systems with no Galilean invariance, in contrast to the their higher dimensional counterparts. A similar effect exists for CDW order as well. At commensurate filling, the boost can open a charge gap for systems of spinless fermions. For spin $1/2$ fermions, a boost applied to a fully gapped system can produce a Luther-Emery fluid with gapped spin and gapless charge excitations.

%%%%%%%%%%%%%%%%%%%%%%%%%%%%%%%%%%%%%%%%%%%%%%%%%%%
%%%% SUPPLEMENTAL SECTION
%%%%-----------------------------------------------
\def\makeSM{1}
\ifdefined\makeSM

%%%%%%%%%%%%%%%%%%%%%%%%%%%%%%%%%%%%%%%%%%%5
\newwrite\tempfile
\immediate\openout\tempfile=junkSM.\jobname
\immediate\write\tempfile{\noexpand{\thepage} }
\immediate\closeout\tempfile

\clearpage

\newpage

%%%%%%%%%%%%%%%%%%%%%%%%%%%%%%%%%%%%%%%%%%%%%%%%%%%%%%%%%%%%%%%%%%%%%%%%%%%%%%%%%%%%%%%%%%%%%%%%%%%

%\end{document}
%UNCOMMENT THIS TO GENERATE FILE WITHOUT SUPPELEMENTAL MATERIAL

\appendix

\renewcommand{\appendixname}{}
\renewcommand{\thesection}{{S\arabic{section}}}
\renewcommand{\theequation}{\thesection.\arabic{equation}}
 
\setcounter{page}{1}
\setcounter{figure}{0}

\widetext

\centerline{\sc Supplemental Material}
\centerline{\sc  for}
\centerline{\bf \savetitle}
\centerline{by}
\centerline{Sayonee Ray,  Subroto Mukerjee and Vijay B. Shenoy}

\section{Modified Luttinger parameters under the effect of the boost}
Under the effect of the boost, the right movers and left movers have different effective Fermi velocities:
\begin{equation}
\begin{split}
v^F_R\ &=\ v^F(u)+w(u) \\
v^F_L\ &=\ v^F(u)-w(u).
\end{split}
\end{equation}

The kinetic part of the Hamiltonian is:
\begin{equation}
H_K\ =\ \int \bigg[ v^F_R \rho_R^\dagger(x) \rho_R(x) + v^F_L \rho_L^\dagger(x) \rho_L(x) \bigg] dx,
\end{equation}
where the $\rho_{R, L}$ are the density operators of the right and left movers.
We can introduce the fields $\phi$ and $\theta$ \cite{giamarchi}, 
\begin{equation} \label{phi-theta}
\begin{split}
\nabla \phi(x)\ &=\ -\pi [ \rho_R(x)+\rho_L(x) ], \\
\nabla \theta(x)\ &=\ \pi [ \rho_R(x)-\rho_L(x) ],
\end{split}
\end{equation}
where $\nabla \phi$ is the $ q \sim 0$ part of the density fluctuation and $\nabla \theta$ is the current operator.
In terms of the bosonic fields, the kinetic part of the Hamiltonian becomes:
\begin{equation}
H_K\ =\ \frac{1}{\pi} \int \bigg[ v^F(u) \bigg( (\nabla\phi (x))^2 + (\nabla\theta (x))^2 \bigg) -2 w(u) (\nabla\phi (x))(\nabla\theta (x)) \bigg] dx.
\end{equation}
For spinless fermions, the relevant interaction processes are the $g_4$ and $g_2$ scattering processes \cite{giamarchi}:
\begin{equation}
\begin{split}
V_{g_4}\ &=\ \frac{g_4}{2} [\rho_R(x) \rho_R(x)+\rho_L(x) \rho_L(x)], \\
V_{g_2}\ &=\ g_2 \rho_R(x) \rho_L(x).
\end{split}
\end{equation}
These interaction processes are not affected by the boost. To see this, we note that they arise from density-density interactions (as also do umklapp terms to be considered later) which preserve the translational symmetry of the lattice. Such interactions are of the form 
\begin{equation}
V \sim \sum_{k_1,k_2,k_3,k_4} V(k_1-k_3) c^\dagger_{k_1} c^\dagger_{k_2}c_{k_3}c_{k_4} \delta(k_1+k_2-k_3-k_4),
\label{Eq:interaction}
\end{equation}
in momentum space,  where $k_1,k_2,k_3$ and $k_4$ are the momenta of the fermions. Since all momenta are fully summed over in the above form and terms in the summation only involve differences of momenta, a boost leaves it unaffected since it adds $u$ to {\em all} momenta.

Using \ref{phi-theta}, the total Hamiltonian with boost is:
\begin{equation} \label{htot}
\mathcal{H}= v (u)\int_{-L/2}^{L/2} dx \left[K (u) :\left(\Pi(x) \right)^2: + \frac{1}{K(u)} :\left(\nabla \phi(x) \right)^2: -2 w(u) :(\nabla\phi (x))(\Pi(x)): \right],
\end{equation}
where \begin{eqnarray}\nonumber
v(u) & = & \sqrt{\left(v^F(u) + \frac{g_4}{2\pi} \right)^2 - \left( \frac{g_2}{2\pi} \right)^2},\\ 
K(u)& = & \sqrt{\frac{v^F(u) - \left(\frac{g_2}{2\pi}-\frac{g_4}{2\pi}\right)}{v^F(u) + \left(\frac{g_2}{2\pi}+\frac{g_4}{2\pi}\right)}}.
\label{Eq:boost_para}
\end{eqnarray}
\vspace{0.5cm}
and, $\Pi(x) = \nabla\theta(x)$.

The Hamiltonian \ref{htot}, can be written as Eq. (\ref{Eq:Ham_tilde}) of the main text, in terms of the new field $\tilde{\phi}$:
\begin{equation}
\mathcal{H}= \frac{v(u)}{2}\int_{-L/2}^{L/2} dx \left[K(u) :\left(\tilde{\Pi}(x) \right)^2: + \frac{1}{K(u)} :\left(\nabla \tilde{\phi}(x) \right)^2: \right],
%\label{Eq:Ham_tilde}
\end{equation}
where:
\begin{eqnarray}\nonumber 
\partial_t \tilde{\phi} & = & \partial_t \phi + w(u)\nabla \phi\\ 
\nabla \tilde {\phi} & = & \nabla \phi,
\label{Eq:new_fields}
\end{eqnarray}
with $\tilde{\Pi}=\partial_t \tilde{\phi}$. Eq. (\ref{Eq:new_fields}) implies 
\begin{equation}
\tilde{\phi}(x,t)=\phi(x+w(u)t,t),
\label{Eq:field_boost}
\end{equation}
Thus, the factor $w(u)$ acts as a velocity in the Galilean transformation of the coordinates $(x,t)$ to obtain the new field $\tilde{\phi}$.

\section{Pairing Susceptibility}
The pairing susceptibility is given by:
\begin{equation} \label{suscp}
\chi_{\rm{pair}}(q=0,\omega)\ =\ \frac{1}{\Omega} \sum _k \frac{ f(\xi _k) - f(-\xi _{-k}) }{\omega - \xi(k) - \xi(-k) + i\delta},
\end{equation}
where $ f(\xi_k)$ is the Fermi distribution at energy $\xi(k)$, where  $\xi(k) = -2 \hopt \cos{k} - \mu$. $\Omega$ is the volume of the system. 

Linearizing the dispersion about the Fermi points, we get~\cite{giamarchi}:
\begin{eqnarray}\nonumber 
\xi(k)& \simeq & v_R^F (k - k_R^F),\ \ k \sim k_R^F \\
\xi(-k)& \simeq & v_L^F (-k - k_L^F),\ \ k \sim k_L^F,
\label{Eq:xi}
\end{eqnarray}
where: $ v^F_s=v^F(u)+s w(u) $ and $s=1(-1)$ for the right(left) movers, as given in the main text. $k_s^F$ is the Fermi momentum form the right($s=R$) and left($s=L$) movers. 

The pairing susceptibility is largest when the Fermi level is in the middle of the band, which corresponds to half-filling. We thus, calculate it as a function of boost for this value of filling using Eq. (\ref{Eq:xi}). The result is shown in Fig.~\ref{Fig:susceptibility}, which is a plot of the ratio of the susceptibility of the boosted system to that of the unboosted system $\chi(u)/\chi(0)$. It can be seen that the susceptibility increases as a function of the boost consistent with the strengthening of superconducting order. This is true even for values of filling different from half-filling. Note that there is a divergent factor of $\log T$, where $T$ is the temperature that cancels between the numerator and denominator of the quantity $\chi(u)/\chi(0)$.

\begin{figure}[]
\centering
\includegraphics[width=8cm, height=6cm]{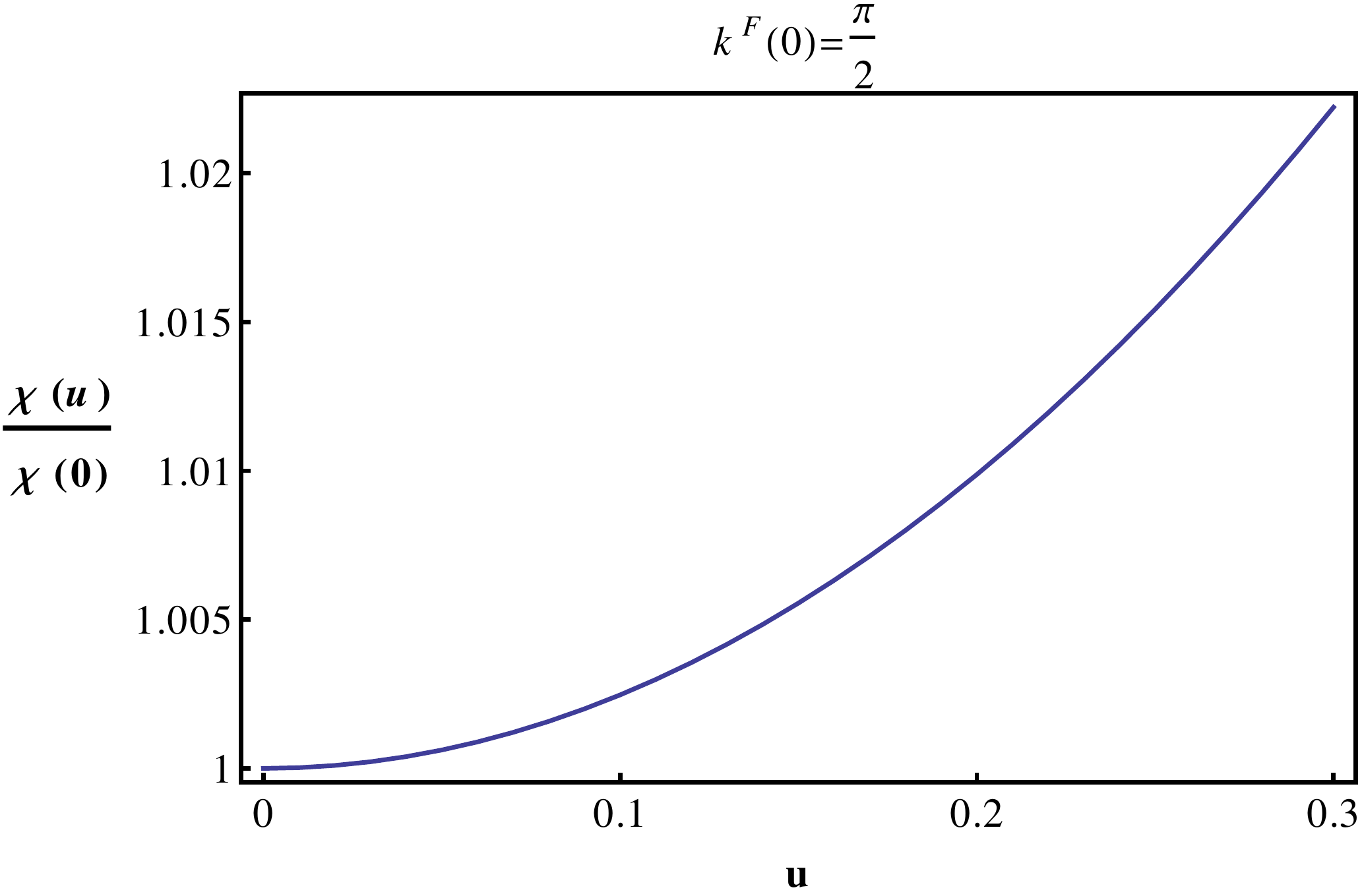} 
\caption{The ratio of the boosted and unboosted pairing susceptibility $\chi(u)/\chi(0)$ at half-filling as a function of $u$. It can be seen that the susceptibility increases with $u$ which is consistent with the strengthening of superconducting order. Note that there is a divergent factor of $\log T$, where $T$ is the temperature that cancels between the numerator and denominator of the quantity $\chi(u)/\chi(0)$.}
\label{Fig:susceptibility}
\end{figure}

\section{Calculation of Correlation functions}
The operator for obtaining the pairing correlations of spinless fermions is : $ O_{SU} (x)\ =\ \psi^{\dagger}(x) \psi^{\dagger}(x^+) $ and that for density wave is : $ O_{CDW} (x)\ =\ \psi^{\dagger}_R(x) \psi_L(x)+ \psi^{\dagger}_L(x) \psi_R(x)$ ,
where $x^+$ is the displacement from x by a lattice parameter and $\psi$ is the single particle operator given by:
\begin{equation} \label{psi}
\psi_s(x)\ =\ \frac{U_r}{2 \pi a} e^{i s k^F_s x} e^{-i[s \phi(x)-\theta(x)]}.
\end{equation}
$s=+(-)1$ denotes right(left) movers, $U_s$ are the Klein factors and $a$ is an ultraviolet cut off.

For $u=0$, the SU and CDW correlation functions are given by~\cite{giamarchi,miranda}:
\begin{eqnarray}\nonumber 
\langle O_{\rm SU}(x,t) O_{\rm SU}^\dagger (x',t') \rangle & \sim & \left( \frac{1}{\sqrt{(x-x')^2+\left[v^F(0) \right]^2 (t-t')^2}} \right)^{1/K(0)}, \\
\langle O_{\rm CDW}(x,t) O_{\rm CDW} (x',t') \rangle & \sim & \cos{[2 \kfo(x-x')]} \left( \frac{1}{\sqrt{(x-x')^2+\left[v^F(0) \right]^2 (t-t')^2}} \right)^{K(0)},
\label{Eq:corr0xt}
\end{eqnarray}

This immediately yields 
\begin{eqnarray} \nonumber
\langle O_{\rm SU}(x) O_{\rm SU}^\dagger (x') \rangle & \sim & \left( \frac{1}{|x-x'|} \right)^{1/K(0)}, \\
\langle O_{\rm CDW}(x) O_{\rm CDW} (x') \rangle & \sim & \cos{[2 \kfo(x-x')]} \left( \frac{1}{|x-x'|} \right)^{K(0)},
\label{Eq:corr0}
\end{eqnarray}
for the equal time correlation functions.

If the transformed field \ref{Eq:field_boost} is used in \ref{htot} and \ref{psi}, following the same procedure that gives Eq. (\ref{Eq:corr0xt}), we obtain the space-time correlation functions for $u \neq 0$:
\begin{eqnarray}\nonumber 
\langle O_{\rm SU}(x,t) O_{\rm SU}^\dagger (x',t') \rangle & \sim & e^{i 2 u(x-x')} \left( \frac{1}{\sqrt{\left[x-x' + w(u)(t-t')\right]^2+\left[v^F(u) \right]^2 (t-t')^2}} \right)^{1/K(u)}, \\
\langle O_{\rm CDW}(x,t) O_{\rm CDW} (x',t') \rangle & \sim & \cos{[2 \kfo(x-x')]} \left( \frac{1}{\sqrt{\left[x-x' + w(u)(t-t')\right]^2+\left[v^F(u) \right]^2 (t-t')^2}} \right)^{K(u)},
\end{eqnarray}
from which it follows that the equal time correlation functions are
\begin{eqnarray} \nonumber
\langle O_{\rm SU}(x) O_{\rm SU}^\dagger (x') \rangle & \sim & e^{i 2 u(x-x')} \left( \frac{1}{|x-x'|} \right)^{1/K(u)}, \\
\langle O_{\rm CDW}(x) O_{\rm CDW} (x') \rangle & \sim & \cos{[2 \kfo(x-x')]} \left( \frac{1}{|x-x'|} \right)^{K(u)},
\label{Eq:corr1}
\end{eqnarray}

\section{RG flow in the presence of umklapp}
Hamiltonian with umklapp:
\begin{equation}
\mathcal{H_\nu}= \frac{v_\nu}{2}\int_{-L/2}^{L/2} dx \left[K_\nu :\left(\tilde{\Pi_\nu}(x) \right)^2: + \frac{1}{K_\nu} :\left(\partial_x \tilde{\phi_\nu}(x) \right)^2:  + \frac{g_\nu}{a^2} :\cos(\alpha_\nu \tilde{\phi_\nu}(x)): \right],
\label{Eq:Ham_umklapp_sup}
\end{equation}
where $\nu=\rho(\sigma)$ for charge(spin) sector.
Following \cite{gnt}, the RG equations are:
\begin{equation}
 \begin{split}
  \frac{d g_\nu}{d l}\ &=\ \bigg(2 - \frac{\alpha_\nu^2 K_\nu}{4 \pi} \bigg) g_\nu \\
  \frac{d K_\nu}{d l}\ &=\ - \frac{K_\nu^3}{2}g_\nu^2 
 \end{split}
\end{equation}

Putting $ K^c_\nu\ =\ \frac{8 \pi}{\alpha_\nu^2} $, define  $ K_\nu = K^c_\nu(1+ \frac{h_\nu}{2}) $ and $g^{\perp}_\nu\ =\ K^c_\nu g_\nu$ . \hphantom{} \\
The RG equations are:
$\frac{d g^{\perp}_\nu}{d l}\ =\ - h_\nu g^{\perp}_\nu $
and $ \frac{d h_\nu}{d l}\ =\ - (g^{\perp}_{\nu})^{2}$ .
The flow equation is given by:
$
 h_\nu^2 - (g^{\perp}_\nu)^2\ =\ (h^{*}_\nu)^{2} .
$
Suppose the unboosted system starts the flow from $h_{\nu,i}(0)$ and $g^\perp_{\nu,i} (0)$, then, the flow follows:
\begin{equation}
 h_{\nu,i}(0)^2 - g^\perp_{\nu,i} (0)^2\ =\ h^{*}_\nu (0)^{2} .
\end{equation}
We need to see where the flow is headed under the effect of the boost, i.e, we need to find $h^{*}_\nu (u)$ .
Then the modifications to the unboosted equations are obtained by introducing the parameters as functions of $u$:
\begin{equation}
 \begin{split}
  K_{\nu,i} (u)\ &=\ K^c_{\nu} \bigg[1+ \frac{h_{\nu,i}(u)}{2}\bigg] ,\\
  g^{\perp}_{\nu,i}(u)\ &=\ g^\perp_{\nu,i} (0) \bigg[ \frac{v_{\nu,i}(0)}{v_{\nu,i}(u)} \bigg] ,
  \label{Eq:renormkg}
 \end{split}
\end{equation}
where  $v_{\nu,i}(0)$ and $v_{\nu,i}(u)$ are the initial values of the renormalized Fermi velocity in the absence of boost and in the presence of boost respectively.
Note that the interaction parameter $V$ in Eq. (\ref{Eq:interaction}) corresponds to the product $v_\nu g_\nu$ in Eq. (\ref{Eq:Ham_umklapp_sup}). $V$ does not change under the effect of the boost which implies that $g_\nu(u) \sim 1/v_\nu(u)$, which yields the second of Eqs. (~\ref{Eq:renormkg}).
Define:
\begin{align}
 a_{\nu,i}(u) &= v^{F}_{i} (u) + \frac{g_{4 \nu,i}}{2 \pi} - \frac{g_{2 \nu,i}}{2 \pi} ,\\
 b_{\nu,i}(u) &= v^{F}_{i} (u) + \frac{g_{4 \nu,i}}{2 \pi} + \frac{g_{2 \nu,i}}{2 \pi} .
\end{align} 
Then, $K_{\nu,i}(u) = \sqrt{\frac{a_{\nu,i}(u)}{b_{\nu,i}(u)}}$ ,
$ v_{\nu,i}(u) = \sqrt{a_{\nu,i}(u) b_{\nu,i}(u)}$ and
$ v^{F}_{i} (u) = v^F_{i}(0) (1+ f(u))$ , where,
$ f(u) = \cos{u} -1$ for the usual tight-binding model,  and $v^F_{i}(0)$ is the initial (bare) value of the Fermi velocity in the absence of boost.

Solving for $K_{\nu,i}(u)$ in terms of $a_{\nu,i}(u)$ and $b_{\nu,i}(u)$ :
\begin{eqnarray} \nonumber
 K_{\nu,i}(u)\ & = & \ \sqrt{\frac{a_{\nu,i}(u)}{b_{\nu,i}(u)}} \\ \nonumber
 & = & \sqrt{\frac{a_{\nu,i}(0) + v^F_{i}(0) f(u)}{b_{\nu,i}(0) + v^F_{i}(0) f(u)}} \\ \nonumber
 & = & \sqrt{\frac{a_{\nu,i}(0)}{b_{\nu,i}(0)} \bigg(1 + \frac{v^F_{i}(0)}{a_{\nu,i}(0)} f(u)\bigg)\bigg(1 - \frac{v^F_{i}(0)}{b_{\nu,i}(0)} f(u) \bigg)} \\ 
 & = & K_{\nu,i}(0) \bigg[1 + \frac{v^F_{i}(0) f(u)}{2 a_{\nu,i}(0) b_{\nu,i}(0)} (b_{\nu,i}(0) - a_{\nu,i}(0)) \bigg] 
\end{eqnarray}

Solving for $v_{\nu,i}(u)$:
\begin{eqnarray} \nonumber
 v_{\nu,i}(u)\ & = & \sqrt{a_{\nu,i}(0) b_{\nu,i}(0)} \\ \nonumber
 & = & \sqrt{[a_{\nu,i}(0) + v^F_{i}(0) f(u)][b_{\nu,i}(0) + v^F_{i}(0) f(u)]} \\
 & = & v_{\nu,i}(0) \bigg(1 + \frac{v^F_{i}(0) (a_{\nu,i}(0) + b_{\nu,i}(0)) }{2 a_{\nu,i}(0) b_{\nu,i}(0)} f(u) \bigg) .
\end{eqnarray}

\vspace{0.5cm}

$K_{\nu,i}(0)$ and $v_{\nu,i}(0)$ are the unboosted value of the initial points from where the flow starts. $K_{\nu,i}(0)$ is then expanded about its critical point $K^c_\nu$ : $ K_{\nu,i} (0)\ =\ K^c_\nu \big(1 + \frac{h_{\nu,i}(0)}{2} \big) $ . 
\vspace{0.5 cm}

Thus,
\begin{eqnarray} \nonumber
 K_{\nu,i}(u)\ &=& K^c_\nu \big(1 + \frac{h_{\nu,i}(0)}{2} \big) \bigg[1 + \frac{v^F_{i}(0) f(u)}{2 a_{\nu,i}(0) b_{\nu,i}(0)} (b_{\nu,i}(0) - a_{\nu,i}(0)) \bigg] \\ \nonumber
 &=& K^c_\nu \bigg[ 1 + \frac{h_{\nu,i}(0)}{2} \bigg(1 + \frac{v^F_{i}(0) f(u)}{2 a_{\nu,i}(0) b_{\nu,i}(0)} (b_{\nu,i}(0) - a_{\nu,i}(0)) \bigg) \\ \nonumber
 & + &  \frac{v^F_{i}(0) f(u)}{2 a_{\nu,i}(0) b_{\nu,i}(0)} (b_{\nu,i}(0) - a_{\nu,i}(0)) \bigg]  \\
 & = & K^c_\nu \left(1+\frac{h_{\nu,i}(u)}{2} \right)
\end{eqnarray}

where \begin{eqnarray} \nonumber
 h_{\nu,i}(u)\ &=& h_{\nu,i}(0) \bigg(1 + \frac{v^F_{i}(0) f(u)}{2 a_{\nu,i}(0) b_{\nu,i}(0)} (b_{\nu,i}(0) - a_{\nu,i}(0)) \bigg) +  \frac{v^F_{i}(0) f(u)}{ a_{\nu,i}(0) b_{\nu,i}(0)} (b_{\nu,i}(0) - a_{\nu,i}(0)) \\
 &=& 2 \bigg( \frac{K_{\nu,i}(0)}{K^c_\nu}-1 \bigg)-\frac{v^F_{i}(0) f(u)}{v_{\nu,i}(0) K^c_\nu} \bigg(K_{\nu,i}(0)^2-1 \bigg) ,
\end{eqnarray}
\vspace{0.5cm}

and, \begin{eqnarray} \nonumber
      g^{\perp}_{\nu,i} (u)\ &=& g^{\perp}_{\nu,i}(0) \bigg( \frac{v_{\nu,i}(0)}{v_{\nu,i}(u)} \bigg) \\ \nonumber
      &=& g^{\perp}_{\nu,i}(0) \bigg(1 - \frac{v^F_{i}(0) (a_{\nu,i}(0) + b_{\nu,i}(0)) }{2 a_{\nu,i}(0) b_{\nu,i}(0)} f(u) \bigg) \\
      &=& g^{\perp}_{\nu,i}(0) \bigg[1- \frac{v^F_{i}(0) f(u)}{2 v_{\nu,i}(0)}\bigg(K_{\nu,i}(0)+\frac{1}{K_{\nu,i}(0)}\bigg) \bigg] . 
     \end{eqnarray}
\vspace{0.5cm}

Using the expressions for $h_{\nu,i}(u)$ and $g^{\perp}_{\nu,i}(u)$ in the flow equation we get the expression for the new fixed point,

\begin{eqnarray} \nonumber
 h_{\nu,i}(u)^2  - g^{\perp}_{\nu,i}(u)^2 \ &=& \bigg[h_{\nu,i}(0) \bigg(1 + \frac{v^F_{i}(0) f(u)}{2 a_{\nu,i}(0) b_{\nu,i}(0)} (b_{\nu,i}(0) - a_{\nu,i}(0)) \bigg) +  \frac{v^F_{i}(0) f(u)}{ a_{\nu,i}(0) b_{\nu,i}(0)} (b_{\nu,i}(0) - a_{\nu,i}(0)) \bigg]^2 \\ \nonumber
  & - & g^{\perp}_{\nu,i}(0)^2 \bigg(1 - \frac{v^F_{i}(0) (a_{\nu,i}(0) + b_{\nu,i}(0)) }{2 a_{\nu,i}(0) b_{\nu,i}(0)} f(u) \bigg)^2 \\ \nonumber
&=& \left[h_\nu^*(0)\right]^2 + \frac{v^F_{i}(0) f(u) }{a_{\nu,i}(0) b_{\nu,i}(0)} \bigg[h_{\nu,i}(0) \bigg(h_{\nu,i}(0) + 2 \bigg)\bigg(b_{\nu,i}(0)-a_{\nu,i}(0)\bigg) \\
& + & g^{\perp}_{\nu,i}(0)^2 \bigg(a_{\nu,i}(0) + b_{\nu,i}(0)\bigg) \bigg] ,
\end{eqnarray}
\vspace{0.5cm}

which is, \begin{equation}
\begin{split}
\left[h_\nu^*(u)\right]^2\ &=\ \left[h_\nu^*(0)\right]^2 + \frac{v^F_{i}(0) f(u) }{a_{\nu,i}(0) b_{\nu,i}(0)} \bigg[h_{\nu,i}(0)  \bigg(h_{\nu,i}(0) + 2 \bigg)\bigg(b_{\nu,i}(0)-a_{\nu,i}(0)\bigg) +  g^{\perp}_{\nu,i}(0)^2 \bigg(a_{\nu,i}(0) + b_{\nu,i}(0)\bigg) \bigg] \\
&=\ \bigg[4 \bigg( \frac{K_{\nu,i}(0)}{K^c_\nu}-1 \bigg)^2 - g^{\perp}_{\nu,i}(0)^2 \bigg] + \frac{v^F_{i}(0) f(u)}{v_{\nu,i}(0) K^c_\nu} \bigg[ 4 \bigg(K_{\nu,i}(0)^2-1 \bigg) \bigg(1-\frac{K_{\nu,i}(0)}{K^c_\nu} \bigg) \\
& + \frac{ g^{\perp}_{\nu,i}(0)^2 K^c_\nu}{K_{\nu,i}(0)} \bigg(K_{\nu,i}(0)^2+1 \bigg) \bigg] ,
\end{split}
\end{equation}
where $h_\nu^*(u)$ gives the new fixed point, as a function of the initial unboosted starting point $K_{\nu,i}(0)$ and $g^{\perp}_{\nu,i}(0)^2$, along with $v_{\nu,i}(0)$, $v^F_{i}(0)$ and the boost $u$ .

\clearpage

\fi %%TO MAKE SUPPLEMENTAL MATERIAL

\end{document}